\documentclass[12pt]{iopart}
\usepackage{iopams}

\usepackage{graphicx}
\usepackage{color}
\usepackage{url}

\newcommand{\bea}{\begin{eqnarray}}
\newcommand{\eea}{\end{eqnarray}}
\newcommand{\beq}{\begin{equation}}
\newcommand{\eeq}{\end{equation}}

\newcommand{\eqref}[1]{(\ref{#1})}

\begin{document}

\def\fun#1#2{\lower3.6pt\vbox{\baselineskip0pt\lineskip.9pt
  \ialign{$\mathsurround=0pt#1\hfil##\hfil$\crcr#2\crcr\sim\crcr}}}
\def\lap{\mathrel{\mathpalette\fun <}}
\def\gap{\mathrel{\mathpalette\fun >}}
\def\kms{{\rm km\ s}^{-1}}
\def\vk{V_{\rm recoil}}


\title[Accuracy of the Post-Newtonian Approximation]
{Note on Accuracy of the Post-Newtonian Approximation
for Extreme-Mass Ratio Inspirals:
Retrograde Orbits}

\author{Ryuichi Fujita}
\address{CENTRA, Departamento de F\'isica, Instituto Superior T\'ecnico,
Universidade de Lisboa, Avenida Rovisco Pais 1, Portugal}

\author{Norichika Sago}
\address{Faculty of Arts and Science, Kyushu University, Fukuoka 819-0395, Japan}

\author{Hiroyuki Nakano}
\address{Faculty of Law, Ryukoku University, Kyoto 612-8577, Japan}
\address{Department of Physics, Kyoto University, Kyoto 606-8502, Japan}

\begin{abstract}

The post-Newtonian approximation is useful to
discuss gravitational waveforms for binary inspirals.
In this note, for retrograde circular orbits
in the equatorial plane,
we discuss the region of validity of
the post-Newtonian approximation by using
the gravitational energy flux derived numerically and analytically
in the black hole perturbation approach.
It is found that
the edge of the allowable region behaves quite intricately,
including prograde orbits
even if we treat higher post-Newtonian orders.
\end{abstract}

\maketitle

\section{Introduction}

The post-Newtonian (PN) approximation
(see, e.g.,~\cite{Blanchet:2013haa,Sasaki:2003xr} for a review)
is widely used not only to derive gravitational waveforms
for binary inspirals, but also to prepare initial parameters
for numerical relativity (NR) simulations
(see, e.g.,~\cite{Baker:2002qf,Husa:2007rh,Campanelli:2008nk}
for quasicircular initial parameters of binary black hole
(BBH) simulations).

In the PN approximation, we assume the slow-motion/weak-field.
To treat the PN approximation, 
it is important to know where the above assumptions are appropriate.
In the case of quasicircular binaries,
the region of validity of the PN approximation is expressed,
for example, by the orbital radius.

Using the PN energy flux for extreme mass ratio inspirals (EMRIs)
in the black hole (BH) perturbation approach,
the region of validity of the PN approximation has been discussed
in~\cite{Poisson:1995vs,Yunes:2008tw,Zhang:2011vha}
(see also~\cite{Mino:1997bx}).
Since we use the assumption of $\mu \ll M$
where the EMRIs are described by a point particle with mass $\mu$
orbiting around a BH with mass $M$, in this approach,
it is possible to calculate the energy flux numerically
and analytically up to very high PN orders.
In our previous paper~\cite{Sago:2016xsp},
we treated the numerical energy flux derived
in~\cite{Fujita:2004rb,Fujita:2009uz}
and the PN flux in~\cite{Fujita:2014eta}
(see also~\cite{Fujita:2011zk,Fujita:2012cm,Sago:2015rpa})
for prograde circular orbits	in the Kerr spacetime,
and analyzed the region of validity.
Although there are some large region of validity for some PN order
(see, e.g., the top panel of Figure 1 in~\cite{Sago:2016xsp}
where the 3PN order ($N=6$) has a larger region of validity
than that of the 3.5PN order ($N=7$)),
we have found that higher PN orders give a larger region of validity
as expected. 
In our previous paper and this note, we note that the region of validity is discussed for the gravitational energy flux, i.e. the time-averaged dissipative piece of the first-order gravitational self-force (GSF). We do not consider the conservative piece of the first-order GSF in this note.

Here, an important circular orbit in the equatorial plane
is the innermost stable circular orbit (ISCO).
The ISCO radius in Boyer-Lindquist coordinates is given as~\cite{Bardeen:1972fi}
\bea
 \frac{r_{\rm ISCO}}{M} &=& 3 + Z_2 \mp \sqrt{(3 - Z_1) (3 + Z_1 + 2 Z_2)} \,,
\label{eq:ISCO_radius}
\eea
where
\bea
 Z_1 &=& 1 + (1 - (a/M)^2)^{1/3} [(1 + a/M)^{1/3} + (1 - a/M)^{1/3} ] \,,
 \cr
 Z_2 &=& \sqrt{3 (a/M)^2 + Z_1^2} \,,
\eea
and the upper and lower signs refer to the prograde (direct)
and retrograde orbits (with the Kerr parameter $0 \leq a \leq M$), respectively.
Given the orbital radius $r$, the orbital velocity is calculated by 
\bea
 v = (M\Omega)^{1/3} \,; \quad
 M \Omega = \frac{M^{3/2}}{r^{3/2}+aM^{1/2}} \,.
 \label{eq:v_r}
\eea
In Table~\ref{tab:ISCO}, the ISCO radius, frequency and velocity
are summarized for various $a/M$.

\begin{table}[!ht]
\caption{The ISCO radius ($r_{\rm ISCO}$),
frequency ($\Omega_{\rm ISCO}$) and velocity ($v_{\rm ISCO}$)
for various Kerr spin parameters ($a$).}
\label{tab:ISCO}
\begin{center}
\begin{tabular}{|cccc|}
\hline
$a/M$ & $r_{\rm ISCO}/M$ & $M\Omega_{\rm ISCO}$ & $v_{\rm ISCO}$ \\
\hline
1.0 (retrograde) & 9.000000000 & 0.03571428571 & 0.3293168780
\\
0.9 (retrograde) & 8.717352279 & 0.03754018063 & 0.3348359801
\\
0.5 (retrograde) & 7.554584713 & 0.04702732522 & 0.3609525320
\\
0.0 & 6.000000000 & 0.06804138173 & 0.4082482904
\\
0.5 (prograde) & 4.233002531 & 0.1085883589 & 0.4770835292
\\
0.9 (prograde) & 2.320883043 & 0.2254417086 & 0.6086179484
\\
1.0 (prograde) & 1.000000000 & 0.5000000000 & 0.7937005260
\\
\hline
\end{tabular}
\end{center}
\end{table}

For the retrograde ISCOs in Table~\ref{tab:ISCO},
we see larger ISCO radii, smaller ISCO frequencies and velocities 
than those for the prograde ISCOs. 
As a naive expectation, 
the region of validity of the PN approximation for the retrograde case
will be smaller than that for the prograde case.
Despite this spin dependence, we usually require
a same frequency for various configuration
to produce hybrid PN-NR waveforms~\cite{Ajith:2012az,Aasi:2014tra}.
Also, we need to search gravitational waves (GWs) with various spin directions
of binary systems
since the plausible spin direction is not known.
Interestingly, a recent GW event, GW170104
suggests that both spins aligned with the orbital angular momentum
of the BBH system are disfavored~\cite{Abbott:2017vtc}.

In our previous paper~\cite{Sago:2016xsp}, we only discussed several 
cases of prograde orbits and did not discuss retrograde orbits since 
we expected that the region of validity in the PN approximation 
for retrograde orbits would behave similar to that for prograde ones. 
Motivated by the observation of GW170104, 
however, we reanalyze the region of validity in the PN approximation 
both for prograde and retrograde orbits more carefully 
and find complicated behaviors in the region of validity which have not been 
noticed in our previous paper (see, e.g., Fig.~\ref{fig:N_vs_AR_summary}). 
We would expect that additional results might be of interest to 
people working on numerical relativity simulations, 
and would like to summarize results in this note.

The note is organized as follows. In Section~\ref{sec:EAR},
we review the definition of the edge of allowable region (AR) briefly
which was introduced in our previous paper~\cite{Sago:2016xsp}.
In Section~\ref{sec:results},
we show the edge of the allowable region from the eccentricity estimation
for retrograde circular orbits in the equatorial plane.
Finally, we summarize and discuss
the results presented in our previous paper and this note in Section~\ref{sec:dis}. 
In this note, we use the geometric unit system, where $G=c=1$.

\section{Edges of the allowable region}\label{sec:EAR}

To discuss the region of validity of the PN approximation,
we use the gravitational energy flux $F_{g}$
for a point particle orbiting in circular orbits around a BH.
The basic equations in the BH perturbation approach
are the Regge-Wheeler-Zerilli~\cite{Regge:1957td,Zerilli:1971wd}
and Teukolsky~\cite{Teukolsky:1973ha} equations.
The energy flux is computed in two ways:
one is the numerical approach which gives the ``exact'' value
in the numerical accuracy, and the other is the analytical approach
where we employ the PN approximation.

In the analytical approach,
the gravitational energy flux is written as
\bea
 \frac{dE^{(N)}}{dt} = - F_{\rm Newt} F^{(N)} \,,
 \label{eq:EF}
\eea
where $F_{\rm Newt}$ is the Newtonian flux,
\bea
 F_{\rm Newt} = \frac{32}{5} \left(\frac{\mu}{M}\right)^2 v^{10} \,,
\eea
and $F^{(N)}$ is the ($N/2$)PN result
normalized by the Newtonian flux.
$v$ is the circular orbital velocity.
We denote the exact value obtained in the numerical approach
as $F$.

In our previous paper~\cite{Sago:2016xsp},
we followed the analyses presented in~\cite{Yunes:2008tw}
(based on~\cite{BenderOrszag})
and obtained consistent results with~\cite{Yunes:2008tw,Zhang:2011vha}.
As an alternative method which is the simplest analysis
and a practical method for the region of validity, 
we have considered an allowance $\delta$ for the difference
between the numerical and analytical results as
\bea
|F-F^{(N)}| < \delta \,.
\eea
Here, when we discuss the quasicircular evolution,
the deviation from the exact energy flux induces the ``artificial'' orbital eccentricity $e$, that shows the deviation from the ``exact'' quasicircular evolution.
Assuming the eccentric orbit with a semimajor axis $r_0$,
\bea
 r(t) \sim r_{0}\left[ 1-e \cos \left(\frac{v t}{r_{0}}\right) \right] \,.
\eea
where we have already known that the Newtonian orbit is sufficient
to estimate the eccentricity~\cite{Sago:2016xsp},
the orbital eccentricity is expressed as
\bea
 e \equiv& \left(\frac{dE}{dr}\right)^{-1}
 \frac{1}{v}\, F_{\rm Newt}\, |F-F^{(N)}|
 \,,
 \label{eq:ee}
\eea
where $E$ is the orbital energy of the particle,
\bea
 \frac{E}{\mu} =&
 {\frac {{r}^{3/2}-2\,M\sqrt {r}+a\sqrt {M}}{{r}^{3/4} \left( {r}^{3/2}
 -3\,M\sqrt {r}+2\,a\sqrt {M} \right)^{1/2} }} \,.
\eea
The relation between the orbital radius and velocity is given
by~\eqref{eq:v_r}.

In NR simulations of BBH~\cite{Pretorius:2005gq, Campanelli:2005dd, Baker:2005vv},
an iterative procedure 
to obtain low-eccentricity initial orbital parameters
is used~\cite{Boyle:2007ft,Pfeiffer:2007yz,Buonanno:2010yk,
Purrer:2012wy,Buchman:2012dw}.
According to~\cite{Buchman:2012dw},
we need a $t = O(1000M)$ numerical evolution
(where $M$ denotes the total mass for comparable mass binaries) 
to reduce the eccentricity by about a factor of 10. 
For example,
we find the eccentricity in the initial parameters
for NR simulations of GW150914~\cite{Abbott:2016blz,TheLIGOScientific:2016qqj,
TheLIGOScientific:2016wfe,TheLIGOScientific:2016uux} is
$\sim 0.0012$ (RIT) and $0.0008$ (SXS)~\cite{Lovelace:2016uwp}
(see also BBH simulation
catalogs~\cite{Mroue:2013xna,Jani:2016wkt,Abbott:2016apu,Healy:2017psd}
and references therein).
Since quasicircular initial parameters are an important input
for NR simulations~\cite{Healy:2017zqj},
we treat the edge of the allowable region by using the eccentricity
to discuss the region of validity of the PN approximation.

\section{Results}\label{sec:results}

\begin{figure}[!ht]
\center
\includegraphics[width=0.8\textwidth]{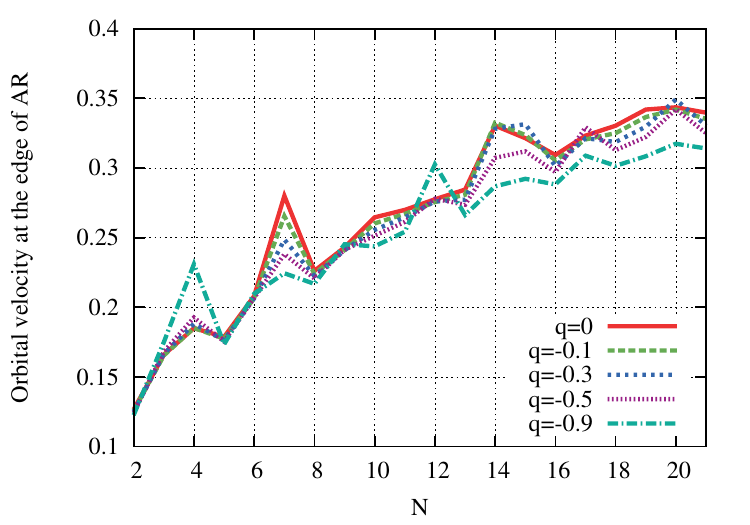}
\\
\includegraphics[width=0.8\textwidth]{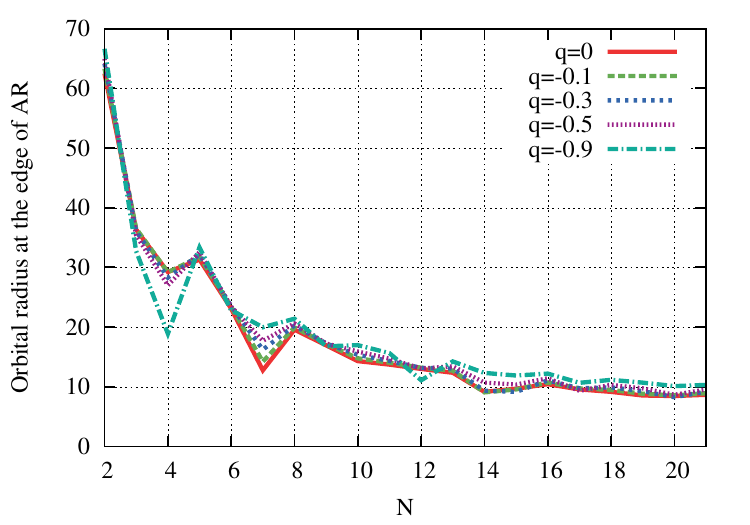}
\caption{
  The edge of the allowable region from the eccentricity estimation
  in the case of $e \leq 1 \times 10^{-5}$
  in terms of the orbital velocity (top) and radius (bottom).}
\label{fig:N_vs_AR_r}
\end{figure}

We set a restriction on $e$ from the error in the energy flux as
\bea
 e \leq 1 \times 10^{-5} \,,
 \label{eq:eerest}
\eea
where this reference value is adopted from
the lowest eccentricity in NR simulations presented in~\cite{Ajith:2012az}.
Combining~\eqref{eq:ee} with~\eqref{eq:eerest}, 
the edge of the allowable region is presented 
in terms of the orbital velocity (top), $v^{(N)}$, and radius (bottom)
for retrograde circular orbits 
in Kerr ($q=a/M=0,\,-0.1,\,-0.3,\,-0.5,\,-0.9$) cases
in Figure~\ref{fig:N_vs_AR_r}.
Here, the (normalized) numerical energy flux $F$
and the (normalized) PN flux $F^{(N)}$ in~\eqref{eq:ee}
are calculated by~\cite{Fujita:2004rb,Fujita:2009uz}
and~\cite{Fujita:2014eta}, respectively.

\begin{figure}[!t]
\center
\includegraphics[width=0.8\textwidth,clip=true]{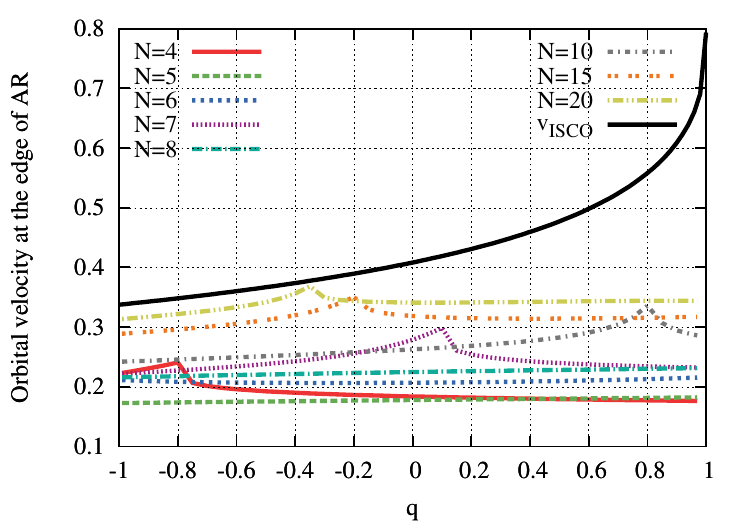}
\\
\includegraphics[width=0.8\textwidth,clip=true]{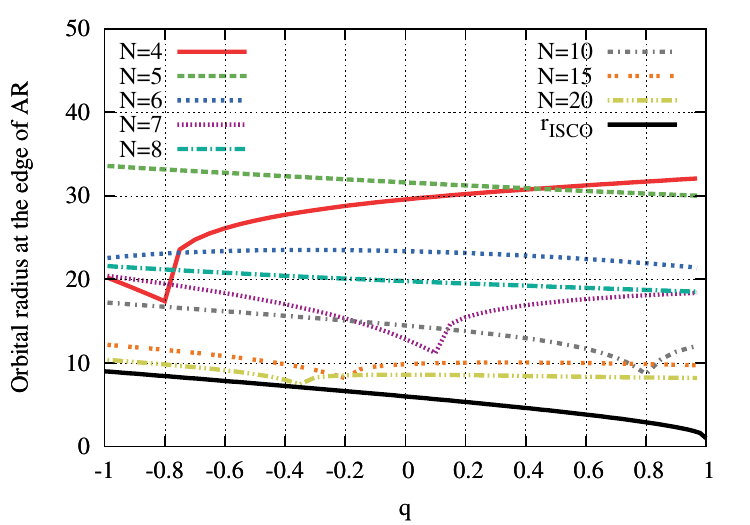}
\caption{
Edges of the allowable region for the orbital velocity $v_{\rm AR}$
(top) and radius $r_{\rm AR}$ (bottom)
for all range of the non-dimensional
Kerr parameter, $q=a/M$.
Here, $q>0$ and $q<0$ are for prograde and retrograde orbits, respectively.}
\label{fig:N_vs_AR_summary}
\end{figure}

At 3.5PN order ($N=7$), we see a clear tendency for
the allowable region
to decrease when $q$ becomes more negative.
Also, in large $N$, 
the allowable region
for large negative $q$ values 
has a similar behavior to the ISCO given in~\eqref{eq:ISCO_radius}
(see also Table~\ref{tab:ISCO} and Figure~\ref{fig:N_vs_AR_summary}),
but it is not so clear.

\section{Summary and Discussions}\label{sec:dis}

In this note, we have used EMRIs for the analysis.
For the current ground-based GW detectors,
Advanced LIGO (aLIGO)~\cite{TheLIGOScientific:2014jea}, 
Advanced Virgo (AdV)~\cite{TheVirgo:2014hva}, 
KAGRA~\cite{Somiya:2011np,Aso:2013eba},
comparable mass ratio binaries are the target.
To extrapolate the result in this note to these binaries
where individual objects have mass $m_1$ and $m_2$,
and nondimensional spin $\chi_1$ and $\chi_2$,
it may be possible to consider $M$, $\mu$ and $q=a/M$ as 
the total mass ($m_1+m_2$), the reduced mass ($m_1m_2/(m_1+m_2)$) of the system,
and an effective spin
$\chi_{\rm eff}=(m_1\chi_1+m_2\chi_2)/(m_1+m_2)$~\cite{Ajith:2009bn}
(see also~\cite{Damour:2001tu}), respectively.

Figure~\ref{fig:N_vs_AR_summary}
shows the edges of the allowable region for the orbital velocity $v_{\rm AR}$
and radius $r_{\rm AR}$
from the eccentricity estimation (\eqref{eq:ee} with~\eqref{eq:eerest})
for all range of Kerr parameter.
The ISCO values given in Table~\ref{tab:ISCO} are also shown.
For example, it is found at 2.5PN and 4PN order ($N=5$ and $8$, respectively)
that the orbital radius at the edge of the allowable region 
for large negative values of $q$ tends to be larger and 
has a similar $q$ dependence to the ISCO radius. 
This suggests that when we start numerical relativity simulations
of binary systems with the 2.5PN and 4PN initial orbital parameters,
we need to consider a larger initial orbital separation
for large negative $q$ cases.

\begin{figure}[!t]
\center
\includegraphics[width=0.8\textwidth,clip=true]{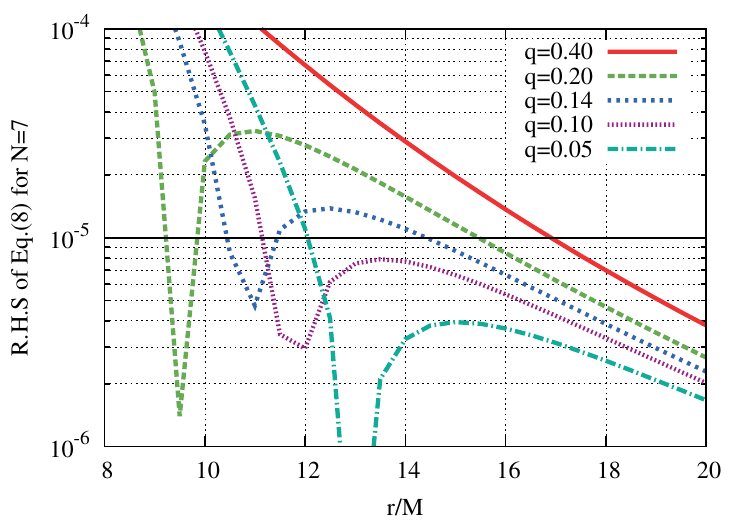}
\\
\includegraphics[width=0.8\textwidth,clip=true]{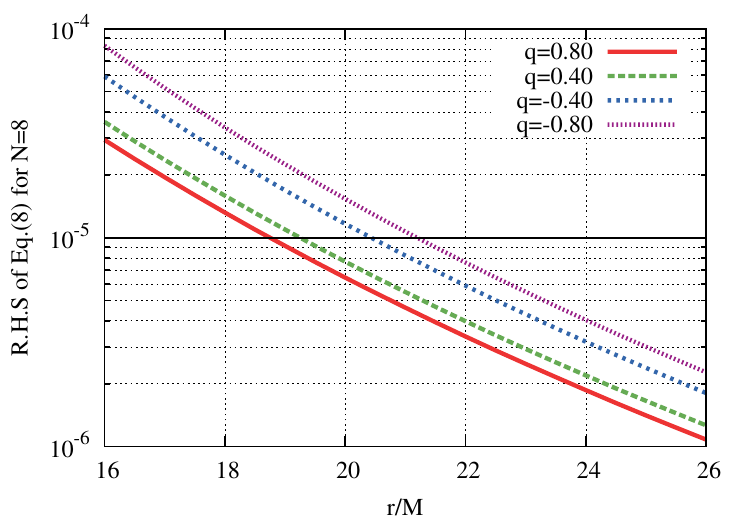}
\caption{
The right hand side of \eqref{eq:ee}
as a function of the orbital radius 
for various non-dimensional Kerr parameters, $q=a/M$
in the case of $N=7$ (top) and $8$ (bottom).}
\label{fig:fe_r_N4}
\end{figure}

On the other hand, at 3.5PN order ($N=7$)
the allowable region basically decreases for large $q$.
This may be a possible reason why 
the initial parameters given by Ref.~\cite{Healy:2017zqj}
for nearly maximally spinning BBH simulations~\cite{Zlochower:2017bbg}
do not work well.
It is noted in Figure~\ref{fig:N_vs_AR_summary} that 
discontinuous change with respect to $q$
arise from different sequences of solutions 
for the equality \eqref{eq:ee} with $e=1 \times 10^{-5}$. 
The top panel of Figure~\ref{fig:fe_r_N4} shows complicated behaviors 
of the right hand side of \eqref{eq:ee} for $N=7$ and various $q$ 
as a function of the orbital radius. 
From this figure, we can find that 
there are several solutions in the orbital radius for $q=0.20$ and $0.14$ 
when we set $e=1 \times 10^{-5}$, 
while there is a single solution for $q=0.40$, $0.10$ and $0.05$. 
In the case that \eqref{eq:ee} has several solutions, we choose the largest
solution in the orbital radius as the edge of the allowable region,
which is larger than the one
that gives the local maximum of the right hand side of \eqref{eq:ee}. 
Then the solutions vary smoothly from $q=0.40$ to $q=0.14$. 
The discontinuous change of solutions in the orbital radius 
with respect to $q$ might appear between $q=0.14$ and $0.10$ 
when we set $e=1 \times 10^{-5}$. 
This is because the local maximum of the right hand side of \eqref{eq:ee} 
for $q=0.10$ becomes smaller than $1 \times 10^{-5}$
and the solution in the orbital radius for $q=0.10$ becomes smaller than 
the one that gives the local minimum of the right hand side of \eqref{eq:ee}. 
The locations of discontinuities depend on the restriction on $e$.
Contrarily, we see simple behaviors for $N=8$ in the bottom panel
of Figure~\ref{fig:fe_r_N4}, and this gives the continuous variation
in Figure~\ref{fig:N_vs_AR_summary}. 
Even if we treat higher PN order, say, $N=20$,
it is difficult to estimate a simple $q$-dependence of the edge
of the allowable region. 
Therefore, the above conclusion derived from the 2.5PN and 4PN analyses
is not held. We should note that
any local peaks in Figures~\ref{fig:N_vs_AR_r} and~\ref{fig:N_vs_AR_summary}
do not indicate the best PN approximation in a given order
as mentioned in~\cite{Sago:2016xsp}.
We always welcome higher PN results 
in order to investigate whether the allowable region becomes larger.

Although the allowable region has been discussed
for the gravitational energy flux, i.e., 
the first order dissipative self-force
in our previous paper~\cite{Sago:2016xsp} and this note,
the post-adiabatic effects of the self-force,
i.e., the first order conservative
and the second order dissipative self-forces should be studied.
As for the Schwarzschild background, 
the radius of convergence of the PN series 
for the first order conservative self-force
has been shown in~\cite{Johnson-McDaniel:2015vva} (see also~\cite{Kavanagh:2015lva}).
For the Kerr case, 
It is possible to discuss the radius of convergence
by using an analytic result presented in~\cite{Kavanagh:2016idg}.
Furthermore,~\cite{Bini:2016dvs} is usable
in the case of eccentric orbits around a Kerr BH.
For future work,
we will analyze the allowable region by using the above results
to connect our analysis with the comparable mass ratio binaries.

\ack

RF's work was funded through H2020 ERC Consolidator Grant
``Matter and strong-field gravity: New frontiers in Einstein's theory''
(MaGRaTh-646597).
This work was also supported by
JSPS Grant-in-Aid for Scientific Research (C), No.~JP16K05356 (NS)
and No.~JP16K05347 (HN),
and MEXT Grant-in-Aid for Scientific Research
on Innovative Areas, ``New developments in astrophysics
through multi-messenger observations of gravitational wave sources'',
No.~JP24103006 (HN). 
Some numerical computations were carried out at the Yukawa Institute Computer Facility. 



\section*{References}

\end{document}